\documentclass[12pt]{article}
\usepackage{graphicx}
\usepackage{amsmath}
\usepackage{verbatim}
\usepackage{mathrsfs}
\usepackage{float}
\usepackage{subfigure}
\usepackage{epsfig}

\begin{document}
\begin{titlepage}
\title{\bf\Large Oblique Corrections in the MSSM at One Loop. I. Scalars  \vspace{18pt}}
\author{\normalsize Yao~Yu and Sibo~Zheng \vspace{12pt}\\
{\it\small  Department of Physics, Chongqing University, Chongqing 401331, P.R. China}\\}

\date{}
\maketitle \voffset -.3in \vskip 1.cm \centerline{\bf Abstract}
\vskip .3cm
This paper is devoted to reconsider the one-loop oblique corrections arising from the scalar superpartners in the MSSM,
i.e, the squarks, the sleptons and the scalars in Higgs sector.
We explicitly present the complete one-loop forms of self-energy corrections to the gauge bosons of SM electroweak gauge groups,
as well as their descendants the $S$, $T$ and $U$ parameters,
which can be directly applied to constrain the parameter space of the MSSM.
Our results about one-loop self energies are found to agree with Ref. \cite{SUSY, Group3}.
Nevertheless, the $S$, $T$ and $U$ parameters aren't in agreement with Ref. \cite{Group2A}.
 \thispagestyle{empty}

\end{titlepage}

\section{Introduction}
With LHC keeping running on searches of the standard model (SM) Higgs with mass around 125 GeV \cite{LHC: Higgs},
we are at the dawn when studies of Higgs physics become a field of the precise test,
and we find out whether low-scale supersymmetry is going to show up.

Among frameworks in which the implication of the recent LHC results in new physics are explored,
the electroweak precise test always stands as an important way for studying phenomenology of Higgs physics.
The logic is that SM (naturally including the Higgs particle $h$) and its extensions such as supersymmetric SM
can be analyzed through considering the quantum corrections,
more concretely oblique corrections \cite{Peskin1, Peskin2} in these theories to the electroweak precise test observables.
In SM the oblique corrections mainly depend on the Higgs particle mass $m_h $ and top quark mass,
it follows that electroweak precise test  provides useful constraint on $m_{h}$ \cite{Group1B}.
For example, it is very successful in analyzing technicolor models \cite{Peskin2}.

This method can  be similarly applied in SUSY models,
from which the masses of supersymmetric particles can be constrained (see, e.g, \cite{Group3}).
But this method is not very efficient due to some reasons.
One is that there are so many mass parameters in SUSY models.
Another is that the bounds on these masses are not yet sufficient before running of the LHC collider \cite{LHC: SUSY}.
So even in the minimal supersymmetric standard model (MSSM)
it is impossible to make firm claims.
In order to extract useful information, as discussed in the literature,
various limits such as highly degenerate \cite{Group1A, Group1B}
or super heavy masses \cite{Group2A, Group2B, Group2C} of supersymmetric particles are considered in SUSY models.

However, these approximations are too simple to be suitable
when the present LHC data is incorporated into the MSSM.
Also,
it can be verified that the analytic results related to our discussion usually do not consistent in the literature.
Due to these observations,
we should ascertain the oblique corrections in the MSSM at first.
In this paper, we will address this issue by using the two-component
formalism of Lagrangian for supersymmetric particles under the electroweak group.
The main outcomes are two-fold.
At first, the results we obtain can be applied to most general cases without taking any assumptions about the quark- and lepton scalar masses.
Moreover, our results can't reduce to those of Ref. \cite{Group2A} when we use the same limit as the authors took.
The difference is probably due to the missing or over counting of Feynman loops involving heavy scalars as the internal lines. 

We want to emphasize that while this work is being prepared,
the LHC results \cite{LHC: Higgs} indicate that its main task has been transformed from \em{discovery} \em  to \em{precise tests} \em  of the SM-like Higgs and \em{discovery} \em of supersymmetry.
The oblique corrections to electroweak observables can still serve as a window to explore the parameter space of superpartners' masses
by combining the data collected by the LHC and other colliders.
The implications to MSSM and NMSSM along this line will be addressed elsewhere \cite{1303.1900}.

The paper is organized as follows.
In section 2 , we derive the bosonic oblique corrections in the MSSM.
We divide the task into three parts,
i.e, the squark sector, the slepton sector, and the Higgs sector sector.
We will consider the fermionic contributions in the further \cite{new}.
In section 3,
we then make preliminary checks on the results presented in section 2 in terms of the fact that the radiative corrections to electroweak
mixing angle is finite.
In section 4, we  present the bosonic results of $S$, $T$ and $U$ parameters in the MSSM.
The property that these parameters are also finite is more stringent examination on the results than the one in the section 3.
In section 5, we conclude and discuss our main results.
We find that our results about one-loop self energies are found to agree with Ref. \cite{SUSY, Group3}.
Nevertheless, the $S$, $T$ and $U$ parameters do not math with those of Ref. \cite{Group2A}.
In appendix A explicit expressions for the functionals related to one-loop graphs are presented.
We would like to emphasize that the on-shell renormalization scheme is used throughout this note.


\section{One-loop Bosonic Contributions}
Since we deal with the analytic calculation of one-loop self energy of SM electroweak gauge bosons,
it is convenient to use the two-component formalism both for the supersymmetric scalars and fermions.

\subsection{Lagrangian for Electroweak Scalar Doublets }
To derive the Feynman rules, we refer to the Lagrangian for scalar doublets $\Phi^{T}=(\phi_{1}, \phi_{2})$ under the representation of electroweak gauge group,
which can be written as,
 \begin{eqnarray}{\label{G1}}
\mathcal{L}_{H}\sim (\overrightarrow{D}_{\nu}\Phi)^{\dag}(\overrightarrow{D}^{\nu}\Phi)=\Phi^{\dag}\overleftarrow{D}_{\nu}\overrightarrow{D}^{\nu}\Phi
\end{eqnarray}
with the definitions,
\begin{eqnarray}{\label{G2}}
\overrightarrow{D}_{\nu}\Phi=\left(\begin{array}{cc}
                  \overrightarrow{\partial}_{\nu}-i(\frac{e}{2sc}-\frac{es}{c}Q_{1}) Z_{\nu}-ieQ_{1}A_{\nu}& -i\frac{e}{\sqrt{2}s}W_{\nu}^{+} \\
                  -i \frac{e}{\sqrt{2}s}W_{\nu}^{-} &\overrightarrow{\partial}_{\nu}+i(\frac{e}{2sc}+\frac{es}{c}Q_{2})Z_{\nu}-ieQ_{2}A_{\nu}
                  \end{array}\right)\Phi
\end{eqnarray}
Here $Q_{1}$ and $Q_{2}$ represent the $Y$-charges of up-type $\phi_{1}$ and down-type $\phi_{2}$ respectively.
And
\begin{eqnarray}{\label{G3}}
\Phi^{\dag} \overleftarrow{D}_{\nu}=\Phi^{\dag}\left(\begin{array}{cc}
                  \overleftarrow{\partial}_{\nu}+i(\frac{e}{2sc}-\frac{es}{c}Q_{1}) Z_{\nu}+ieQ_{1}A_{\nu}& i\frac{e}{\sqrt{2}s}W_{\nu}^{+} \\
                  i \frac{e}{\sqrt{2}s}W_{\nu}^{-} &\overleftarrow{\partial}_{\nu}-i(\frac{e}{2sc}+\frac{es}{c}Q_{2})Z_{\nu}+ieQ_{2}A_{\nu}
                  \end{array}\right)
\end{eqnarray}
Parameters $s=\sin\theta_{W}$ and $c=\cos\theta_{W}$.

For EW singlet scalars $S$ such as right-hand squarks and sleptons,
the Lagrangian is given by,
 \begin{eqnarray}{\label{G4}}
\mathcal{L}_{H}\sim (D_{\nu}S)^{\dag}(D^{\nu}S)=S^{*}\left(\overleftarrow{\partial}_{\nu}-i\frac{s}{c}eQZ_{\nu}+ieQA_{\nu}\right)\left(\overrightarrow{\partial}^{\nu}+i\frac{s}{c}eQZ^{\nu}-ieQA^{\nu}\right)S
\end{eqnarray}
Here $Q$ denotes the $Y$-charge of $S$ field.

From \eqref{G1} to \eqref{G4}, we can derive the Feynman rules for MSSM scalars (see, e.g,\cite{Hunter}).
In what follows, we take the ' t Hooft-Feynman gauge for non-abelian gauge fields involved.
It turns out that there are three type of graphs as shown in Fig.1 needed to be considered,
\begin{figure}[h!]
\centering
\begin{minipage}[b]{0.8\textwidth}
\centering
\includegraphics[width=5in]{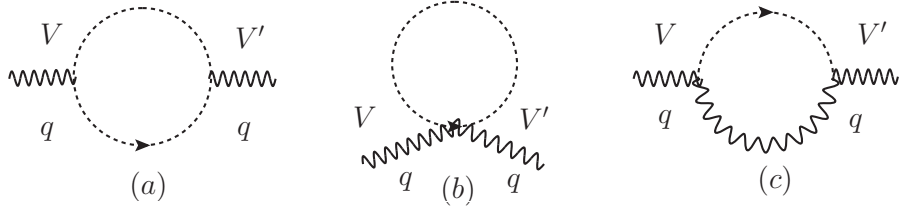}
\end{minipage}%
\caption{Graphs that contribute to the self-energy of SM gauge bosons due to scalar superpartners.}
\end{figure}

\subsection{Squark Sector }
The squark contributions are composed of those coming from three-generation  left-hand squarks
$\tilde{q}_{L i}=(\tilde{u}_{i},\tilde{d}_{ i})$ and their right-hand $\tilde{u}_{R i}$, $\tilde{d}_{R i}$.
Accoring to the Lagrangian \eqref{G1} we find that there are two types of Feynman diagrams that contribute to the one-loop
self-energy of vector bosons, as shown in fig 1. $(a)$  and fig. 1 $(b)$\footnote{
In the next subsection there will be an extra Feynman diagram needed to be counted due to the Higgs VEVs.}.
We find \footnote{We have neglected $1/16$ factor in each $\Pi^{VV'}$ in this section,
which will be restored in the section 4.}
 \begin{eqnarray}{\label{sq1}}
\Pi^{\gamma\gamma}_{L}(q^{2})&=&N_{c} \frac{e^{2}}{9\pi^{2}}\sum_{i=1,2,3}\left[-16A(q^{2}; m^{2}_{\tilde{u}_{i}}, m^{2}_{\tilde{u}_{i}})-4A(q^{2}; m^{2}_{\tilde{d}_{i}}, m^{2}_{\tilde{d}_{i}})+8 a(m^{2}_{\tilde{u}_{i}})+2a(m^{2}_{\tilde{d}_{i}})\right]\nonumber\\
\Pi^{ZZ}_{L}(p^{2})&=&N_{c} \frac{1}{\pi^{2}}\sum_{i=1,2,3}\left[-4\left(\frac{e}{2sc}-\frac{2es}{3c}\right)^{2}A(q^{2}; m^{2}_{\tilde{u}_{i}}, m^{2}_{\tilde{u}_{i}})-4\left(\frac{e}{2sc}-\frac{es}{3c}\right)^{2}A(q^{2}; m^{2}_{\tilde{d}_{i}}, m^{2}_{\tilde{d}_{i}})\right.\nonumber\\
&+&\left.2\left(\frac{e}{2sc}-\frac{2es}{3c}\right)^{2}a( m^{2}_{\tilde{u}_{i}})+2\left(\frac{e}{2sc}-\frac{es}{3c}\right)^{2}a(m^{2}_{\tilde{d}_{i}})\right]
\end{eqnarray}
and
 \begin{eqnarray}{\label{sq0}}
\Pi^{\gamma Z}_{L}(q^{2})&=&N_{c} \frac{e}{3\pi^{2}}\sum_{i=1,2.3}\left[-8\left(\frac{e}{2sc}-\frac{2es}{3c}\right)A(q^{2}; m^{2}_{\tilde{u}_{i}}, m^{2}_{\tilde{u}_{i}})-4\left(\frac{e}{2sc}-\frac{es}{3c}\right)A(q^{2}; m^{2}_{\tilde{d}_{i}}, m^{2}_{\tilde{d}_{i}})\right.\nonumber\\
&+&\left.4\left(\frac{e}{2sc}-\frac{2es}{3c}\right)a(m^{2}_{\tilde{u}_{i}})+4\left(\frac{e}{2sc}-\frac{es}{3c}\right)a(m^{2}_{\tilde{d}_{i}})\right]\nonumber\\
\Pi^{WW}_{L}(q^{2})&=&N_{c} \frac{1}{2\pi^{2}}\frac{e^{2}}{s^{2}}\sum_{i=1,2.3}\left[\left(a(m^{2}_{\tilde{u}_{i}})+a(m^{2}_{\tilde{d}_{i}})\right)
-4A(q^{2}; m^{2}_{\tilde{u}_{i}}, m^{2}_{\tilde{d}_{i}})\right]
\end{eqnarray}
for left-hand squark doublets, and
\begin{eqnarray}{\label{sq2}}
\Pi^{\gamma\gamma}_{R}(q^{2})&=&N_{c}  \frac{e^{2}}{9\pi^{2}}\sum_{i=1,2,3}\left[-16A(q^{2}; m^{2}_{\tilde{u}_{Ri}}, m^{2}_{\tilde{u}_{Ri}})-4A(q^{2}; m^{2}_{\tilde{d}_{Ri}}, m^{2}_{\tilde{d}_{Ri}})+8a( m^{2}_{\tilde{u}_{Ri}})+2a( m^{2}_{\tilde{d}_{Ri}})\right]\nonumber\\
\Pi^{ZZ}_{R}(p^{2})&=&N_{c} \frac{e^{2}}{9\pi^{2}}\frac{s^{2}}{c^{2}}\sum_{i=1,2,3}\left[-16A(q^{2}; m^{2}_{\tilde{u}_{Ri}}, m^{2}_{\tilde{u}_{Ri}})-4A(q^{2}; m^{2}_{\tilde{d}_{Ri}}, m^{2}_{\tilde{d}_{Ri}})+8a( m^{2}_{\tilde{u}_{Ri}})\right.\nonumber \\
&+&\left.2a( m^{2}_{\tilde{d}_{Ri}})\right]
\end{eqnarray}
together with
 \begin{eqnarray}{\label{sq3}}
\Pi^{\gamma Z}_{R}(q^{2})&=&N_{c} \frac{e^{2}}{9\pi^{2}}\frac{s}{c}\sum_{i=1,2.3}\left[16A(q^{2}; m^{2}_{\tilde{u}_{Ri}}, m^{2}_{\tilde{u}_{Ri}})+4A(q^{2}; m^{2}_{\tilde{d}_{Ri}}, m^{2}_{\tilde{d}_{Ri}})-8a(m^{2}_{\tilde{u}_{Ri}})-2a(m^{2}_{\tilde{d}_{Ri}})\right]\nonumber\\
\Pi^{WW}_{R}(q^{2})&=&0
\end{eqnarray}
for right-hand squarks.
Here $A(p^{2}; x, y)$ and $a(x)$ are one-loop integral functions, which are defined in appendix A.
Note that the right-hand sfermions $\tilde{u}_{R i}$ , $\tilde{d}_{R i}$ and $\tilde{e}_{R i}$
have no contributions to the self-energy of W boson.
Also we want to mention that there are four-point vertexes for W bosons from the Lagrangian \eqref{G1},
however they contribute at two-loop.

The effects on $\Pi$ due to FCNC can be taken into account
through the reexpression of squark and slepton sectors with their corresponding mass eigenstates.
For example, the mixing in  mass matrix of $\tilde{u}_{L }$ and $\tilde{u}_{R}$ can be diagonalized via a
unitary matrix $U(\alpha_{1})$
\begin{eqnarray}{\label{E22}}
\left(\begin{array}{c}
                  \tilde{u}_{L} \\
                   \tilde{u}_{R}
                  \end{array}\right)
= \left(\begin{array}{cc}
                  \cos\alpha_{1} & -\sin\alpha_{1} \\
                  \sin\alpha_{1} & \cos\alpha_{1}
                  \end{array}\right)
\left(\begin{array}{c}
                  \tilde{u}_{1} \\
                   \tilde{u}_{2}
                  \end{array}\right)
\end{eqnarray}
such that the vertexes for sfermions coupled to vector bosons can be read from the mass eigenstates $\tilde{u}_{1,2}$.

\subsection{Slepton Sector }
In this sector, the contributions to self-energy of SM vector bosons stem from the
 three-generation left-hand sleptons $\tilde{l}_{Li}=(\tilde{\nu}_{ i}, \tilde{e}_{ i})$ and right-hand $\tilde{e}_{R i}$.
In terms of the general expression in \eqref{G1}, we derive that,
 \begin{eqnarray}{\label{slepton}}
\Pi^{\gamma\gamma}(q^{2})&=& \frac{e^{2}}{\pi^{2}}\sum_{i=1,2,3}\left[-4A(q^{2}; m^{2}_{\tilde{e}_{i}}, m^{2}_{\tilde{e}_{i}})-4A(q^{2}; m^{2}_{\tilde{e}_{Ri}}, m^{2}_{\tilde{e}_{Ri}})+2 a(m^{2}_{\tilde{e}_{i}})+2 a(m^{2}_{\tilde{e}_{R}})\right]\nonumber\\
\Pi^{ZZ}(p^{2})&=&\frac{e^{2}}{\pi^{2}}\sum_{i=1,2,3}\left[\frac{(1-2s^{2})^{2}}{s^{2}c^{2}}\left(-A(q^{2}; m^{2}_{\tilde{e}_{i}}, m^{2}_{\tilde{e}_{i}})+\frac{1}{2}a(m^{2}_{\tilde{e}_{i}})\right)+\frac{1}{2s^{2}c^{2}}a(m^{2}_{\tilde{\nu}_{i}})\right.\nonumber\\
&-&\left.\frac{s^{2}}{c^{2}}\left(4A(q^{2}; m^{2}_{\tilde{e}_{Ri}}, m^{2}_{\tilde{e}_{Ri}})-2a(m^{2}_{\tilde{e}_{Ri}})\right)-\frac{s^{2}}{c^{2}}4A(q^{2}; m^{2}_{\tilde{{\nu}_{i}}},m^{2}_{\tilde{{\nu}_{i}}})\right]\nonumber\\
\Pi^{\gamma Z}(q^{2})&=&\frac{e^{2}}{\pi^{2}}\sum_{i=1,2,3}\left[\frac{1-2s^{2}}{sc}\left(-2A(q^{2}; m^{2}_{\tilde{e}_{i}}, m^{2}_{\tilde{e}_{i}})+a(m^{2}_{\tilde{e}_{i}})\right)+\frac{s}{c}\left(4A(q^{2}; m^{2}_{\tilde{e}_{Ri}}, m^{2}_{\tilde{e}_{Ri}})-2a(m^{2}_{\tilde{e}_{Ri}})\right)\right]\nonumber\\
\Pi^{WW}(q^{2})&=&\frac{1}{2\pi^{2}}\frac{e^{2}}{s^{2}}\sum_{i=1,2.3}\left[\left(a(m^{2}_{\tilde{\nu}_{i}})+a(m^{2}_{\tilde{e}_{i}})\right)
-4A(q^{2}; m^{2}_{\tilde{\nu}_{i}}, m^{2}_{\tilde{e}_{i}})\right]
\end{eqnarray}
As we have mentioned in the previous subsection,
there is no contribution to $\Pi^{WW}$ from the right-hand $\tilde{e}_{R i}$.
The discussion about taking the masses mixings among left- and right-hand sclars is similar to that about \eqref{E22} via intorducing a unitary matrix $U(\alpha_{2})$
\begin{eqnarray}
\left(\begin{array}{c}
                  \tilde{e}_{L} \\
                   \tilde{e}_{R}
                  \end{array}\right)
= \left(\begin{array}{cc}
                  \cos\alpha_{2} & -\sin\alpha_{2} \\
                  \sin\alpha_{2} & \cos\alpha_{2}
                  \end{array}\right)
\left(\begin{array}{c}
                  \tilde{e}_{1} \\
                   \tilde{e}_{2}
                  \end{array}\right)
\end{eqnarray}
So far we have dealt with evaluating the first two graphs in fig. 1,
now we proceed to discuss the Higgs sector in which the last graph has to be included.

\subsection{Higgs sector}
The calculation of contributions coming from the neutral $A^{0}$, $H^{0}$
as well as the charged real scalar $H^{\pm}$ is similar to those of sfermion sector but more involved.
One reason is that there is an extra Feynman diagram fig. 1$(c)$ needed to be considered due to the new couplings with the Higgs VEVs $v_{\mu}$ and $v_{d}$.
The other reason is that the expressions when we transform from Higgs gauge eigenstates to their mass eigenstates are complicated,
 \begin{eqnarray}{\label{E23}}
H_{\mu}=\left(\begin{array}{c}
                  \sin\beta~G^{+} +\cos\beta~H^{+} \\
                  \frac{1}{\sqrt{2}}v_{\mu}+\frac{1}{\sqrt{2}}\cos\alpha~h^{0}+\frac{1}{\sqrt{2}}\sin\alpha~H^{0}+\frac{i}{\sqrt{2}}\sin\beta~G^{0}+\frac{i}{\sqrt{2}}\cos\beta~A^{0}
                  \end{array}\right)
\end{eqnarray}
and
 \begin{eqnarray}{\label{E24}}
H_{d}=\left(\begin{array}{c}
                 \frac{1}{\sqrt{2}} v_{d}-\frac{1}{\sqrt{2}}\sin\alpha~h^{0}+\frac{1}{\sqrt{2}}\cos\alpha~H^{0}-\frac{i}{\sqrt{2}}\cos\beta~G^{0}+\frac{i}{\sqrt{2}}\sin\beta~A^{0}  \\
                -\cos\beta~G^{-} +\sin\beta~H^{-}
                  \end{array}\right)
\end{eqnarray}
Here $G^{0}$ and $G^{\pm}$ are Nambu-Goldstone modes under the 't Hooft-Feynman gauge.
$\tan\beta=v_{\mu}/v_{d}$ and angle $\alpha$ is introduced in the unitary matrix
so as to  diagonalize the mass matrix of scalars.
The mass spectra for scalars in the Higgs sector can be explicitly found in \cite{Martin97}.

Evaluate the Feynman diagrams gives rise to,
\begin{eqnarray}{\label{H1}}
\Pi^{\gamma\gamma}(q^{2})&=& \frac{e^{2}}{\pi^{2}}\left[-4A(q^{2}; m^{2}_{G^{+}}, m^{2}_{G^{+}})-4A(q^{2}; m^{2}_{H^{+}}, m^{2}_{H^{+}})\right.\nonumber\\
&+&\left. 2a(m^{2}_{G^{+}})+2a(m^{2}_{H^{+}})\right.\nonumber\\
&+&\left. \frac{\pi^{2}e^{2}}{2s^{2}}\left(\sin\beta v_{\mu}+\cos\beta v_{d}\right)^{2}b_{0}(q^{2}; m^{2}_{W}, m^{2}_{G^{+}})\right]\nonumber\\
\Pi^{ZZ}(q^{2})&=&\frac{e^{2}}{4\pi^{2}s^{2}c^{2}}\left\{a(m^{2}_{h^{0}})+a(m^{2}_{H^{0}})+a(m^{2}_{G^{0}})+a(m^{2}_{A^{0}})
\right.\nonumber\\
&+&\left. 2(1-2s^{2})^{2}\left[a(m^{2}_{G^{+}})+a(m^{2}_{H^{+}})\right]\right\}\nonumber\\
&+&\frac{e^{4}}{4\pi^{2}s^{4}c^{4}}\left[(\cos\alpha~v_{\mu}-\sin\alpha~v_{d})^{2}b_{0}(q^{2}; m^{2}_{Z}, m^{2}_{h^{0}})
\right.\nonumber\\
&+&\left.
(\sin\alpha~v_{\mu}+\cos\alpha~v_{d})^{2}b_{0}(q^{2}; m^{2}_{Z}, m^{2}_{H^{0}})\right.\nonumber\\
&+&\left.2s^{4}c^{2}(\sin\beta~v_{\mu}+\cos\beta~v_{d})^{2}b_{0}(q^{2}; m^{2}_{W}, m^{2}_{G^{+}})\right]\nonumber\\
&-&\frac{e^{2}}{\pi^{2}s^{2}c^{2}}\left\{(1-2s^{2})^{2}\left[A(q^{2}; m^{2}_{G^{+}}, m^{2}_{G^{+}})+A(q^{2}; m^{2}_{H^{+}}, m^{2}_{H^{+}})\right]\right.\nonumber\\
&+&\left.\cos^{2}(\alpha-\beta)\left[A(q^{2}; m^{2}_{h^{0}}, m^{2}_{A^{0}}+A(q^{2}; m^{2}_{H^{0}}, m^{2}_{G^{0}}))\right]
\right.\nonumber\\
&+&\left. \sin^{2}(\alpha-\beta)\left[A(q^{2}; m^{2}_{h^{0}}, m^{2}_{G^{0}})+A(q^{2}; m^{2}_{H^{0}}, m^{2}_{A^{0}})\right]\right\}
\end{eqnarray}
and
\begin{eqnarray}{\label{H2}}
\Pi^{\gamma Z}(q^{2})&=&-\frac{e^{2}(1-2s^{2})}{2\pi^{2}sc}\left[2a(m^{2}_{G^{+}})+2a(m^{2}_{H^{+}})\right.\nonumber\\
&+&\left. 4A(q^{2}; m^{2}_{G^{+}}, m^{2}_{G^{+}})-4A(q^{2}; m^{2}_{H^{+}}, m^{2}_{H^{+}})\right]\nonumber\\
&-&\frac{e^{4}}{2\pi^{2}sc}(\sin\beta v_{\mu}+\cos\beta v_{d})^{2}b_{0}(q^{2}; m^{2}_{W}, m^{2}_{G^{+}})\nonumber\\
\Pi^{WW}(q^{2})&=&\frac{e^{2}}{4\pi^{2}s^{2}}\left[2a(m^{2}_{H^{+}})+2a(m^{2}_{G^{+}})+a(m^{2}_{H^{0}})+
a(m^{2}_{A^{0}})+a(m^{2}_{G^{0}})+a(m^{2}_{h^{0}})\right]\nonumber\\
&+&\frac{e^{4}}{4\pi^{2}s^{2}c^{2}}(\sin\beta v_{\mu}+\cos\beta v_{d})^{2}\left[c^{2}b_{0}(q^{2}; 0, m^{2}_{G^{+}})+s^{2}b_{0}(q^{2}; m^{2}_{Z}, m^{2}_{G^{+}})\right]\nonumber\\
&+&\frac{e^{4}}{4\pi^{2}s^{4}}\left[(\cos\alpha v_{\mu}-\sin\alpha v_{d})^{2}b_{0}(q^{2}; m^{2}_{W}, m^{2}_{h^{0}})\right.\\
&+&\left.(\sin\alpha v_{\mu}+\cos\alpha v_{d})^{2}b_{0}(q^{2}; m^{2}_{W}, m^{2}_{H^{0}})\right]\nonumber\\
&-&\frac{e^{2}}{\pi^{2}s^{2}}\left\{\sin^{2}(\beta-\alpha)\left[A(q^{2}; m^{2}_{h_{0}}, m^{2}_{G^{+}})+A(q^{2}; m^{2}_{H_{0}}, m^{2}_{H^{+}})\right]\right.\nonumber\\
&+&\left.\cos^{2}(\alpha+\beta)\left[A(q^{2}; m^{2}_{h_{0}}, m^{2}_{H^{+}})
+A(q^{2}; m^{2}_{H_{0}}, m^{2}_{G^{+}})\right]\right.\nonumber\\
&+&\left.A(q^{2}; m^{2}_{A_{0}}, m^{2}_{H^{+}})+A(q^{2}; m^{2}_{G_{0}}, m^{2}_{G^{+}})\right\}\nonumber
\end{eqnarray}
In terms of the corrections arising from the SM ,
the superpartners' contributions in MSSM can be separated from \eqref{H1} and \eqref{H2} as,
\begin{eqnarray}{\label{H3}}
\Pi^{\gamma\gamma}(q^{2})&=& \frac{e^{2}}{\pi^{2}}\left[-4A(q^{2}; m^{2}_{H^{+}}, m^{2}_{H^{+}})+2a(m^{2}_{H^{+}})\right]\nonumber\\
\Pi^{ZZ}(q^{2})&=&\frac{e^{2}}{4\pi^{2}s^{2}c^{2}}\left[a(m^{2}_{H^{0}})+2a(m^{2}_{A^{0}})
+2(1-2s^{2})^{2}a(m^{2}_{H^{+}})\right]\nonumber\\
&+&\frac{e^{2}}{\pi^{2}s^{2}c^{4}}m^{2}_{W}\cos^{2}(\alpha-\beta)\left[b_{0}(q^{2}; m^{2}_{Z}, m^{2}_{H^{0}})-b_{0}(q^{2}; m^{2}_{Z}, m^{2}_{H^{0}}\right]\nonumber\\
&-&\frac{e^{2}}{\pi^{2}s^{2}c^{2}}\left\{(1-2s^{2})^{2}A(q^{2}; m^{2}_{H^{+}}, m^{2}_{H^{+}})\right.\nonumber\\
&+&\left.\cos^{2}(\alpha-\beta)[A(q^{2}; m^{2}_{h^{0}}, m^{2}_{A^{0}})+A(q^{2}; m^{2}_{H^{0}}, m^{2}_{G^{0}})]\right.\nonumber\\
&-&\left. \cos^{2}(\alpha-\beta)A(q^{2}; m^{2}_{h^{0}}, m^{2}_{G^{0}})+\sin^{2}(\alpha-\beta)A(q^{2}; m^{2}_{H^{0}}, m^{2}_{A^{0}})\right\}
\nonumber\\
\Pi^{\gamma Z}(q^{2})&=&-\frac{e^{2}(1-2s^{2})}{\pi^{2}sc}\left[-2A(q^{2}; m^{2}_{H^{+}}, m^{2}_{H^{+}})+a(m^{2}_{H^{+}})\right]\\
\Pi^{WW}(q^{2})&=&\frac{e^{2}}{4\pi^{2}s^{2}}\left[2a(m^{2}_{H^{+}})+a(m^{2}_{H^{0}})+
a(m^{2}_{A^{0}})\right]\nonumber\\
&+&\frac{e^{2}}{\pi^{2}s^{2}}\cos^{2}(\alpha-\beta)m^{2}_{W}\left[b_{0}(q^{2}; m^{2}_{W}, m^{2}_{H^{0}})-b_{0}(q^{2}; m^{2}_{W}, m^{2}_{h^{0}})\right]\nonumber\\
&+&\frac{e^{2}}{\pi^{2}s^{2}}\cos^{2}(\alpha-\beta)\left[A(q^{2}; m^{2}_{h_{0}}, m^{2}_{G^{+}})-A(q^{2}; m^{2}_{h_{0}}, m^{2}_{H^{+}})
-A(q^{2}; m^{2}_{H_{0}}, m^{2}_{G^{+}})\right]\nonumber\\
&-&\frac{e^{2}}{\pi^{2}s^{2}}\left[\sin^{2}(\alpha-\beta)A(q^{2}; m^{2}_{H_{0}}, m^{2}_{H^{+}})+A(q^{2}; m^{2}_{A_{0}}, m^{2}_{H^{+}})\right]\nonumber
\end{eqnarray}


\section{Preliminary Checks on The Results}
In electroweak theory,
there are various obeservables that can be precisely measured,
however, they depend only on three parameters
which are composed of the gauge coupling constants parameters $g$, $g'$ and VEV $v$ associated with  the scale of spontaneous symmetry breaking.
For example, one can define the weak mixing angle as
 \begin{eqnarray}{\label{C1}}
s^{2}_{W}\equiv1-\frac{m_{W}^{2}}{m_{Z}^{2}}
\end{eqnarray}
Another definition uses,
 \begin{eqnarray}{\label{C2}}
s^{2}_{*}\equiv\frac{g^{' 2}}{g^{2}+g^{' 2}}
\end{eqnarray}
Alternatively, we do this via the accurately known weak-interaction obeservables $\alpha$, $G_F$ and $m_Z$,
\begin{eqnarray}{\label{C3}}
\sin2\theta_{0}\equiv\left(\frac{4\pi \alpha(m_{Z}^{2})}{\sqrt{2}G_{F}m_{Z}^{2}}\right)^{1/2}=0.2307 \pm 0.0005
\end{eqnarray}
which gives us an accurate standard of reference.
The corrections to $\alpha$, $G_F$ and $m_Z$  due to quantum effects of new particle states beyond SM will lead to deviation from the reference point.
So the measured value can be used to constrain the content of these new particles and estimate the bounds of their masses.

Now we use the weak mixing angle as an example to check our results presented in the previous section.
The rational is as follows.
The quantum corrections to observables such as $m_Z$, $m_W$, $\alpha$ and  $G_F$
in low-energy  electroweak theory include corrections to the vector boson propagator, vertex and box  corrections.
The last two types are usually proportional to the ratio of masses of external (light) fermions over masses of heavy particle states.
With the limit that these light masses of  SM fermions are set to zero,
only the first type is important practically.
This type of correction is known as oblique correction,
as they enter the low-energy weak interactions only indirectly.
The different definitions \eqref{C1} to \eqref{C3} all agree at zero order but receive different radiative corrections.
However, the differences, for instance \cite{PeskinBook}
 \begin{eqnarray}{\label{C3}}
s^{2}_{W}-s^{2}_{*}\equiv-\frac{c^{2}}{m_{W}^{2}}\left[\Pi^{WW}(m^{2}_{W})-c^{2}\Pi^{ZZ}(m^{2}_{Z})\right]+\frac{sc^{3}}{m_{W}^{2}}\Pi^{\gamma Z}(m^{2}_{Z})
\end{eqnarray}
due to the radiative corrections must be finite,  and are free of ultraviolet divergence.

For the part of slepton sector, substituting \eqref{slepton} into \eqref{C3} gives,
 \begin{eqnarray}{\label{C4}}
s^{2}_{W}-s^{2}_{*}&=&\frac{c^{2}e^{2}}{\pi^{2}s^{2}m^{2}_{W}}\left[-(1-2s^{2})A(m^{2}_{Z}; m^{2}_{\tilde{e}_{i}},m^{2}_{\tilde{e}_{i}} )-A(m^{2}_{Z}; m^{2}_{\tilde{\nu}_{i}},m^{2}_{\tilde{\nu}_{i}} )\right.\nonumber\\
&+&\left.2A(m^{2}_{W}; m^{2}_{\tilde{e}_{i}},m^{2}_{\tilde{\nu}_{i}} )-s^{2}a(m^{2}_{\tilde{e}_{i}})\right]
\end{eqnarray}
for each generation of left-hand sleptons,
which then gives us the part of divergence is proportional to
 \begin{eqnarray}{\label{C5}}
&-&(1-2s^{2})\left(\frac{1}{2}m^{2}_{\tilde{e}_{i}}-\frac{1}{12}m^{2}_{Z}\right)\nonumber\\
&-&\left(\frac{1}{2}m^{2}_{\tilde{\nu}_{i}}-\frac{1}{12}m^{2}_{Z}\right)\nonumber\\
&+&2\left(\frac{1}{4}m^{2}_{\tilde{e}_{i}}+\frac{1}{4}m^{2}_{\tilde{\nu}_{i}}-\frac{1}{12}m^{2}_{W}-s^{2}m^{2}_{\tilde{e}_{i}}\right)=0\nonumber
\end{eqnarray}
For each generation of right-hand slepton, we find that the contribution to the deference in \eqref{C3} vanishes.

Now we proceed to examine the results in the squark sector.
Substituting \eqref{sq1}  into \eqref{C3} yields,
 \begin{eqnarray}{\label{C6}}
s^{2}_{W}-s^{2}_{*}&=&\frac{e^{2}c^{2}}{\pi^{2}s^{2}m^{2}_{W}}
\left[-(1-\frac{3}{4}s^{2})A(m^{2}_{Z}; m^{2}_{\tilde{u}_{i}},m^{2}_{\tilde{u}_{i}} )-\frac{2}{3}s^{2}a(m^{2}_{\tilde{u}_{i}})
-\frac{1}{3}s^{2}a(m^{2}_{\tilde{d}_{i}})\right.\nonumber\\
&-&\left. (1-\frac{2}{3}s^{2})A(m^{2}_{Z}; m^{2}_{\tilde{d}_{i}},m^{2}_{\tilde{d}_{i}} )+2A(m^{2}_{W}; m^{2}_{\tilde{u}_{i}},m^{2}_{\tilde{d}_{i}} )\right]
\end{eqnarray}
for the left-hand squarks,
from which the part of divergence is proportional to
 \begin{eqnarray}{\label{C7}}
&-&(1-\frac{3}{4}s^{2})(\frac{1}{2}m^{2}_{\tilde{u}_{i}}-\frac{1}{12}m^{2}_{Z})-\frac{2}{3}s^{2}m^{2}_{\tilde{u}_{i}}
-\frac{1}{3}s^{2}m^{2}_{\tilde{d}_{i}}\nonumber\\
&-&(1-\frac{2}{3}s^{2})(\frac{1}{2}m^{2}_{\tilde{d}_{i}}-\frac{1}{12}m^{2}_{Z})
+2(\frac{1}{4}m^{2}_{\tilde{u}_{i}}+\frac{1}{4}m^{2}_{\tilde{d}_{i}}-\frac{1}{12}m^{2}_{Z})=0
\end{eqnarray}

For the part of Higgs sector, substituting \eqref{H1} and \eqref{H2} into \eqref{C3} gives that
the divergent parts in \eqref{C3} are composed of those arising from $a(x)$, $A(x)$ and $b_{0}$ functions.
It turns out the contribution from $a(x)$ function is proportional to
$-\frac{e^{2}c^{2}}{\pi^{2}m^{2}_{W}}(m^{2}_{G^{+}}+m^{2}_{H^{+}})$,
which exactly cancels the part from $A$ function.
And the divergent parts in $b_{0}$ function cancel themselves.


\section{$S$, $T$ and $U$ in the MSSM}
In this section, we derive a set of parameters which estimate the oblique corrections to precise electroweak observables.
These parameters are known as $S$, $T$ and $U$ in the literature \cite{Peskin1, Peskin2}.
As a byproduct we also use the property that these parameters are finite values to examine the results shown in the section 2,
independently of what we have done in the section 3.
The dependence of  $S$, $T$ and $U$ parameters on  $\Pi^{IJ}(p^{2})$ can be perturbatively expanded in powers of the external  momentum squared.
It turns out that these corrections are quite simple \cite{Peskin1, Peskin2},
 \begin{eqnarray}{\label{OB2}}
S&\equiv&-\frac{16\pi}{e^{2}}sc\left[sc\Pi^{\gamma\gamma '}(0)-sc\Pi^{ZZ '}(0)+(c^{2}-s^{2})\Pi^{\gamma Z '}(0)\right]\nonumber\\
T&\equiv&\frac{4\pi}{e^{2}}\left[\frac{\Pi^{WW}(0)}{m^{2}_{W}}-\frac{\Pi^{ZZ}(0)}{m^{2}_{Z}}-\frac{2s}{c}\frac{\Pi^{\gamma Z}(0)}{m^{2}_{Z}}\right]\\
U&\equiv&\frac{16\pi s^{2}}{e^{2}}\left[\Pi^{WW '}(0)-c^{2}\Pi^{ZZ '}(0)-s^{2}\Pi^{\gamma\gamma '}(0)-2c
s\Pi^{\gamma Z '}(0)\right]\nonumber
\end{eqnarray}
where
 \begin{eqnarray}{\label{OB3}}
\Pi^{IJ'}(0)=d^{2}\Pi^{IJ}/dp^{2}\mid_{p^{2}=0}
\end{eqnarray}
$\Pi^{IJ}$ is the part with metric as the coefficient in $\Pi_{\mu\nu}^{IJ}=g_{\mu\nu}\Pi^{IJ}+\cdots$.

In this paper, we consider the calculations of the oblique corrections in MSSM,
which are composed of four parts given a specific vacuum polarization,
\begin{eqnarray}{\label{OB4}}
\Pi_{SUSY}=\Pi_{\tilde{S}}+\Pi_{\tilde{l}}+\Pi_{H}+\Pi_{NC}
\end{eqnarray}
We derive the first three parts in \eqref{OB4} in this paper, and will explore last part for the future \cite{new}.

Firstly,  the slepton sector gives
\begin{eqnarray}{\label{OB5}}
S_{\tilde{l}}&=& \frac{1}{12\pi}\sum_{i=1}^{3}\ln \frac{m^{2}_{\tilde{\nu}_{i}}}{m^{2}_{\tilde{e}_{i}}}\nonumber\\
T_{\tilde{l}}&=&\frac{1}{16\pi s^{2}m^{2}_{W}}\sum_{i=1}^{3}\left[m^{2}_{\tilde{\nu}_{i}}+m^{2}_{\tilde{e}_{i}}-\frac{2m^{2}_{\tilde{e}_{i}}m^{2}_{\tilde{\nu}_{i}}\ln(\frac{m^{2}_{\tilde{\nu}_{i}}}{m^{2}_{\tilde{e}_{i}}})}{m^{2}_{\tilde{\nu}_{i}}-m^{2}_{\tilde{e}_{i}}} \right]\\
U_{\tilde{l}}&=&\frac{1}{\pi}\sum_{i=1}^{3}\left[\frac{m^{2}_{\tilde{e}_{i}}}{3}b'_{0}(0, m^{2}_{\tilde{e}_{i}}, m^{2}_{\tilde{e}_{i}}) -\frac{m^{2}_{\tilde{e}_{i}}+m^{2}_{\tilde{\nu}_{i}}}{6}b'_{0}(0, m^{2}_{\tilde{e}_{i}}, m^{2}_{\tilde{\nu}_{i}})\right.\nonumber\\
&+&\left. \frac{(m^{2}_{\tilde{e}_{i}}-m^{2}_{\tilde{\nu}_{i}})^{2}}{24}b''_{0}(0, m^{2}_{\tilde{e}_{i}}, m^{2}_{\tilde{\nu}_{i}})+\frac{1}{12}f(m^{2}_{\tilde{e}_{i}}, m^{2}_{\tilde{\nu}_{i}})-\frac{1}{12}f(m^{2}_{\tilde{e}_{i}}, m^{2}_{\tilde{e}_{i}})\right]\nonumber
\end{eqnarray}
for the squark sector, we have
\begin{eqnarray}{\label{OB6}}
S_{\tilde{S}}&=& \frac{1}{12\pi}\sum_{i=1}^{3}\ln \frac{m^{2}_{\tilde{d}_{i}}}{m^{2}_{\tilde{u}_{i}}}\nonumber\\
T_{\tilde{S}}&=&\frac{3}{16\pi s^{2}m^{2}_{W}}\sum_{i=1}^{3}\left[m^{2}_{\tilde{u}_{i}}+m^{2}_{\tilde{d}_{i}}-\frac{2m^{2}_{\tilde{u}_{i}}m^{2}_{\tilde{d}_{i}}\ln(\frac{m^{2}_{\tilde{u}_{i}}}{m^{2}_{\tilde{d}_{i}}})}{m^{2}_{\tilde{u}_{i}}-m^{2}_{\tilde{d}_{i}}} \right]\\
U_{\tilde{S}}&=&\sum_{i=1}^{3}\frac{1}{\pi}\left[m^{2}_{\tilde{u}_{i}}b'_{0}(0, m^{2}_{\tilde{u}_{i}}, m^{2}_{\tilde{u}_{i}})+
m^{2}_{\tilde{d}_{i}}b'_{0}(0, m^{2}_{\tilde{d}_{i}},m^{2}_{\tilde{d}_{i}})
-(m^{2}_{\tilde{u}_{i}}+m^{2}_{\tilde{d}_{i}})b'_{0}(0, m^{2}_{\tilde{u}_{i}},m^{2}_{\tilde{d}_{i}})\right.\nonumber\\
&+&\left.\frac{1}{2}f(m^{2}_{\tilde{u}_{i}}, m^{2}_{\tilde{d}_{i}})-\frac{1}{4}f(m^{2}_{\tilde{u}_{i}}, m^{2}_{\tilde{u}_{i}})
-\frac{1}{4}f(m^{2}_{\tilde{d}_{i}}, m^{2}_{\tilde{d}_{i}})\right]\nonumber
\end{eqnarray}
with the finite function $f(x,y)$ is defined as,
 \begin{eqnarray}{\label{OB7}}
f(x,y)=\frac{1}{x-y}\left(x\ln x-y\ln y\right)
\end{eqnarray}
Note that the descents of function $b_{0}$, $b'_{0}$ and $b''_{0}$ in \eqref{OB5} to \eqref{OB6} are given in the appendix B,
which are also finite.

Finally, the Higgs sector yields,
\begin{eqnarray}{\label{OB8}}
\pi S_{H}&=&h(m^{2}_{H^{+}},m^{2}_{H^{+}})
-\sin^{2}(\alpha-\beta)h(m^{2}_{A_{0}}, m^{2}_{H_{0}})\nonumber\\
&+&\cos^{2}(\alpha-\beta)\left[
h(m^{2}_{h_{0}}, m^{2}_{G_{0}})-h(m^{2}_{H_{0}}, m^{2}_{G_{0}})-h(m^{2}_{h_{0}}, m^{2}_{A_{0}})\right]\nonumber\\
&-&\frac{m^{2}_{W}}{c^{2}}\cos^{2}(\alpha-\beta)\left[b'_{0}(0; m^{2}_{Z}, m^{2}_{h_{0}})-b'_{0}(0; m^{2}_{Z}, m^{2}_{H_{0}})\right]
\end{eqnarray}
and \begin{eqnarray}{\label{OB9}}
\pi T_{H}&=&\frac{1}{4s^{2}m^{2}_{W}}\left\{m^{2}_{H^{+}}\ln m^{2}_{H^{+}}+\cos^{2}(\alpha-\beta)\left[m^{2}_{W}(f(m^{2}_{W}, m^{2}_{H_{0}})
-f(m^{2}_{W}, m^{2}_{h_{0}}))\right.\right.\nonumber\\&+&\left.\left.m^{2}_{Z}(f(m^{2}_{Z}, m^{2}_{h_{0}})-f(m^{2}_{Z}, m^{2}_{H_{0}}))\right]\right.\nonumber\\
&+&\left.\left. \cos^{2}(\alpha-\beta)\left[g(m^{2}_{h_{0}}, m^{2}_{G^{+}})+g(m^{2}_{h_{0}}, m^{2}_{A^{0}})+g(m^{2}_{H_{0}}, m^{2}_{G^{0}})
\right.\right.\right.\nonumber\\
&-&\left.\left. g(m^{2}_{h_{0}}, m^{2}_{H^{+}})-g(m^{2}_{H_{0}}, m^{2}_{G^{+}})-g(m^{2}_{h_{0}}, m^{2}_{G^{0}})\right]\right.\nonumber\\
&-&\left.\sin^{2}(\alpha-\beta)g(m^{2}_{H_{0}}, m^{2}_{H^{+}})+\sin^{2}(\alpha-\beta)g(m^{2}_{A_{0}}, m^{2}_{H^{0}})-g(m^{2}_{A_{0}}, m^{2}_{H^{0}})\right\}
\end{eqnarray}
together with
\begin{eqnarray}{\label{OB10}}
\pi U_{H}&=&
h(m^{2}_{H_{+}}, m^{2}_{H^{+}})-h(m^{2}_{A_{0}}, m^{2}_{H^{0}})\nonumber\\
&-&\cos^{2}(\alpha-\beta)m^{2}_{W}
\left[b'_{0} (0; m^{2}_{W}, m^{2}_{H_{0}})-b'_{0} (0; m^{2}_{W}, m^{2}_{h_{0}})\right]\nonumber\\
&+&\cos^{2}(\alpha-\beta)m^{2}_{Z}
\left[b'_{0} (0; m^{2}_{Z}, m^{2}_{h_{0}})-b'_{0} (0; m^{2}_{Z}, m^{2}_{H_{0}})\right]\nonumber\\
&+&\cos^{2}(\alpha-\beta)\left[h(m^{2}_{h_{0}}, m^{2}_{G^{+}})-h(m^{2}_{h_{0}}, m^{2}_{H^{+}})-h(m^{2}_{H_{0}}, m^{2}_{G^{+}})\right.\nonumber\\
&+&\left.h(m^{2}_{h_{0}}, m^{2}_{A^{0}})-h(m^{2}_{h_{0}}, m^{2}_{G^{0}})+h(m^{2}_{G_{0}}, m^{2}_{H^{0}})\right]\nonumber\\
&-&\sin^{2}(\alpha-\beta)\left[h(m^{2}_{H_{0}}, m^{2}_{H^{+}})-h(m^{2}_{A_{0}}, m^{2}_{H^{0}})\right]
\end{eqnarray}
with the finite function $g(x,y)$ and $h(x,y)$ defined as,
 \begin{eqnarray}{\label{OB11}}
g(x,y)&=&\frac{1}{12}(x\ln x+y\ln y)+\frac{x+y}{6}[-1+f(x,y)]+\frac{1}{24}\left(x+y-\frac{2xy}{x-y}\ln \frac{x}{y}\right)\nonumber\\
h(x,y)&=&\frac{1}{6}(x+y)b'_{0}(0; x, y)-\frac{1}{24}(x-y)^{2}b''_{0}(0; x, y)-\frac{1}{12}f(x, y)
\end{eqnarray}
Note that the combination $\cos(\alpha-\beta) $ is given by,
\begin{eqnarray}{\label{OB12}}
\cos^{2}(\alpha-\beta)=\frac{m^{2}_{h}}{m^{2}_{A}}\frac{m^{2}_{Z}-m^{2}_{h}}{m^{2}_{Z}+m^{2}_{A}-2m^{2}_{h}}
\end{eqnarray}

Here a few comments are in order for the results \eqref{OB5} to \eqref{OB10}.
At first, it is obvious that each part in MSSM contributes to the finite $S$, $T$ and $U$ values as required.
There are some constants terms in the expression of $S$ ,$T$ and $U$ parameters,
which naively violate the property that large superpartner masses (compared with $m_{Z}$) lead to the conclusion that the theory decouples from the SM.
But expanding the relevant functions we have defined implies that these constants are cancelled and  the conclusion is restored.
Second, from \eqref{OB8} to \eqref{OB10} it seems that the SM contributions has not been
separated from the MSSM contributions due to some terms involved with SM fields.
But it is crucial to notice that these terms are multiplied by a factor $\cos^{2}(\alpha-\beta)$, as given by \eqref{OB12}.
Under the limit that the superpartners decouple from SM, which results in $\alpha=\beta$,
the results in \eqref{OB8} to \eqref{OB10} reduce to pure MSSM ones.
Finally, unlike in the case of SM, where the sensitivity of these parameters to the SM Higgs mass $m_h$ is logarithmic,
the results in \eqref{OB8} to  \eqref{OB11} demonstrate that this is not strictly true in the MSSM anymore.
According to the simulations about $S$, $T$, $U$ shown in Ref. \cite{1303.1900}, 
the sensitivity of these parameters to the mass of Higgs boson can be either stronger or weaker, 
which depends on choices of parameters left in the Higgs sector. 
Overall, the discrepancy is not so significant as one expects roughly.

Under the assumption that the  breaking of $SU(2)$ symmetry is weak, i.e,
$\Delta_{q_{i}}=m^{2}_{\tilde{e}_{i}}-m^{2}_{\tilde{\nu}_{i}}<<m^{2}_{\tilde{\nu}_{i}}, m^{2}_{\tilde{e}_{i}} $ and
$\Delta_{S_{i}}=m^{2}_{\tilde{d}_{i}}-m^{2}_{\tilde{u}_{i}}<<m^{2}_{\tilde{u}_{i}}, m^{2}_{\tilde{d}_{i}} $,
 we find
 \begin{eqnarray}{\label{OB13}}
S_{\tilde{q}}&\rightarrow&\frac{1}{12\pi}\sum_{i=1}^{3}\frac{\Delta_{S_{i}}}{m^{2}_{\tilde{d}_{i}}}\nonumber\\
T_{\tilde{q}}&\rightarrow&-\frac{3}{16\pi s^{2}}\sum_{i=1}^{3}\frac{\Delta_{S_{i}}}{m^{2}_{W}}\\
U_{\tilde{q}}&\rightarrow&- \frac{1}{4\pi}\sum_{i=1}^{3}\frac{\Delta_{S_{i}}}{m^{2}_{\tilde{d}_{i}}}\nonumber
\end{eqnarray}
and
 \begin{eqnarray}{\label{OB14}}
S_{\tilde{l}}&\rightarrow&-\frac{1}{12\pi}\sum_{i=1}^{3}\frac{\Delta_{q_{i}}}{m^{2}_{\tilde{e}_{i}}}\nonumber\\
T_{\tilde{l}}&\rightarrow&-\frac{1}{16\pi s^{2}}\sum_{i=1}^{3}\frac{\Delta_{q_{i}}}{m^{2}_{W}}\\
U_{\tilde{l}}&\rightarrow&- \frac{1}{24\pi}\sum_{i=1}^{3}\frac{\Delta_{q_{i}}}{m^{2}_{\tilde{e}_{i}}}\nonumber
\end{eqnarray}
From \eqref{OB13} and \eqref{OB14}
one sees that  both in the squark and slepton sector the relative ratio  $|S/U|$ is approximatively  around the unity,
while their values relative to  $T$  depend on the ratios of  $m^{2}/m^{2}_{\tilde{d}_{i}}$ and $m^{2}/m^{2}_{\tilde{e}_{i}}$.

\section{Discussions and Conclusions}
In this paper we have revisited the oblique corrections in the context of the MSSM.
the motivation for exploring these contributions is quite clear since they are useful to
interpret the latest LHC data about the Higgs mass and bounds on superpartner masses
\footnote{Note that there is no need to consider the gauge invariance of these contributions once again,
as this problem in the SM contributions has been properly treated \cite{gauge}. }.
Thus, we reconsider the theoretic calculations under 't Hooft-Feynman gauge,
with the bosonic part as the first step towards the complete answer.

The results presented in this paper are checked by two examinations.
In one examination,
we directly confront our results to the finite radiative correction to
the deference $s^{2}_{W}-s^{2}_{*}$  in each sector.
It shows that the individual contribution deriving from each sector indeed respects this property.
The other examination is by using the property that the $S$, $T$ and $U$ parameters are finite.
We also verify the expectation.

In summary,
our results about one-loop self energies are found to \em{exactly}\em~ agree with Ref. \cite{SUSY, Group3}
\footnote{
The formulae presented in \cite{SUSY} take the SM part into account, however,
do not include the contribution arising from the pure gauge part.
As we missed this point,
we made the claims about disagreement with \cite{SUSY}
in the previous version of this manuscript.}.
Nevertheless, the $S$, $T$ and $U$ parameters do not match with those of Ref. \cite{Group2A},
which can not be explained by the possible difference of renormalization scheme performed.
In particular, our results do not take the scalar mass mixing among left and right hand into account,
therefore correspond to the simple case with vanishing mass mixing.
Take this difference into account,
compare our results with Ref.\cite{Group2A},
one finds that
\begin{itemize}
\item The $S$ and $U$ parameter arising from squark and slepton sector do not agree.
\item The $T$ parameter in \eqref{OB5} and \eqref{OB6} agree with  \cite{Group2A}.
\end{itemize}
We leave the study of oblique corrections arising from the fermions, i.e,
the neutralinos  and charginos elsewhere \cite{new}.\\

~~~~~~~~~~~~~~~~~~~~~~~~~~~~~~~~~~~~~
~~~$\bf{Acknowledgement}$\\
SZ would like to thank Prof. Ken-ichi Hikasa for correspondence.
The work is supported in part by National Natural Science Foundation of China with grant No.11247031.\\

~~~~~~~~~~~~~~~~~~~~~~~~~~~~~~~~~~~~~
~~~$\bf{Conflict~of~Interests}$\\
As the authors of the manuscript,
we do not have a direct financial relation with any commercial
identity mentioned in our paper that might lead to a conflict of
interest for any of the authors.

\appendix
\section{One-loop Integrals}
Two-point funtions $A(p^{2}; x, y)$, $a (x)$ and $b_{0}(p^{2}; x, y)$ are defined as the integrals \cite{Pokorski, functions},
 \begin{eqnarray}{\label{A1}}
\mu^{\epsilon}\int \frac{d^{n}k}{(2\pi)^{n}}\frac{i}{k^{2}-m^{2}}&=&\frac{1}{(4\pi)^{2}}a(m^{2})\nonumber\\
\mu^{\epsilon}\int \frac{d^{n}k}{(2\pi)^{n}}\frac{ik_{\mu}k_{\nu}}
{[k^{2}-m_{1}^{2}][(k+p)^{2}-m^{2}_{2}]}&=&\frac{1}{(4\pi)^{2}}\left[g_{\mu\nu}A(p^{2}; m_{1}^{2}, m_{2}^{2})+\cdots \right]\nonumber\\
\mu^{\epsilon}\int \frac{d^{n}k}{(2\pi)^{n}}\frac{i}
{[k^{2}-m_{1}^{2}][(k+p)^{2}-m^{2}_{2}]}&=&\frac{1}{(4\pi)^{2}}b_{0}(p^{2}; m_{1}^{2}, m_{2}^{2})
\end{eqnarray}
respectively, with $p^2$ is the external line vector boson momentum squared .
Here we have ignored terms that are irrelevant for discussions in the definition of $A(p^{2}; x, y)$.
Explicitly,
 \begin{eqnarray}{\label{A2}}
a(x)&=&-\eta x+a_{F}(x)\nonumber \\
b_{0}(p^{2}; x, y)&=&-\eta+b_{F}(p^{2}; x, y)
\end{eqnarray}
where the divergent factor in these functions are carried by $\eta$,
\begin{eqnarray}{\label{A3}}
 \eta=\frac{1}{\epsilon}+\ln(4\pi)-\gamma_{E}
 \end{eqnarray}
Here $d=4-2\epsilon$, and $\mu$ is the RG scale.
The finite parts $a_F$ and $b_F$ in \eqref{A2} are given by,
 \begin{eqnarray}{\label{A4}}
a_{F}(x)&=&x\left(-1+\ln\frac{x}{\mu^{2}}\right)\nonumber\\
b_{F}(p^{2}; x, y)&=&\int^{1}_{0}dt \ln\frac{tx+(1-t)y-t(1-t)p^{2}}{\mu^{2}}
\end{eqnarray}
respectively.  And
 \begin{eqnarray}{\label{A0}}
A(p^{2}; x, y)&=&\left(\frac{p^{2}}{12}-\frac{x+y}{4}\right)\eta+\frac{1}{12}[a_{F}(x)+a_{F}(y)]
+\frac{1}{6}\left(x+y-\frac{p^{2}}{2}\right)b_{F}(p^{2}; x, y)\nonumber\\
&+&\frac{x-y}{12p^{2}}\left[a_{F}(x)-a_{F}(y)-(x-y)b_{F}(p^{2}; x, y)\right]-\frac{1}{6}\left(x+y-\frac{p^{2}}{3}\right)
\end{eqnarray}
Note that the $A(p^{2}; x, y)$ and $b_{0}(p^{2}; x, y)$ functions are symmetric in $x$ and $y$.

It is also useful to notice some descents of $b_0$ functions
\begin{eqnarray}{\label{A10}}
b^{'}_{0}(0; x, y)&=&-\frac{1}{2}\frac{x+y}{(x-y)^{2}}+\frac{xy}{(x-y)^{3}}\ln\frac{x}{y}\nonumber\\
b^{''}_{0}(0; x, y)&=&\frac{x^{3}+9x^{2}y-9xy^{2}-6xy(x+y)\ln\frac{x}{y}-y^{3}}{3(x-y)^{5}}
\end{eqnarray}

\end{document}